\documentclass{optica-article}

\journal{opticajournal} % for journals or Optica Open

\articletype{Research Article}

\usepackage{lineno}
\begin{document}

\title{Low-Noise Cascaded Frequency Conversion of $637.2$ nm Light to the Telecommunication C-Band in a Single-Waveguide Device}

\author{Fabrice von Chamier,\authormark{1,*} Joscha Hanel,\authormark{1} Chris M\"uller,\authormark{1} Wanrong Li,\authormark{1} Roger Alfredo K\"ogler,\authormark{1} and Oliver Benson\authormark{1}}

\address{\authormark{1}Humboldt-Universität zu Berlin, Institut für Physik, Newtonstraße 15, 12489, Berlin, Germany}

\email{\authormark{*}vonfabri@physik.hu-berlin.de} %% email address is required; see note below about the corresponding author designation

% use {asbstract*} to suppress the copyright line. Copyright information will be added in production

\begin{abstract*} 
Interconnected quantum devices are the building blocks of quantum networks, where state transduction plays a central role. The frequency conversion of photons into the telecommunication C-band is decisive in taking advantage of current low-loss transmission lines. Here, we report the difference frequency conversion of $637.2$ nm fluorescent light from a cluster of NV centers in diamond to tunable wavelengths between $1559.6$ nm and $1565.2$ nm. In order to avoid detrimental noise from spontaneous emissions, we use a two-step conversion device based on a single-pumped periodic poled lithium niobate waveguide. We observed a total external (internal) conversion efficiency of $3.0\pm0.1$ ($20.5\pm0.8$) $\%$ with a noise rate of $2.4\pm0.8$ ($16\pm5$) cps/GHz.
\end{abstract*}

%%%%%%%%%%%%%%%%%%%%%%%%%%  body  %%%%%%%%%%%%%%%%%%%%%%%%%%
\section{Introduction}

Nitrogen-vacancy (NV) centers are suitable candidates for diverse quantum technology applications, ranging between single-photon emission \cite{kurtsiefer2000stable}, spin-photon entanglement \cite{togan2010quantum} and quantum state storage \cite{fuchs2011quantum}. The interconnection of such a platform to disparate sources or to different nodes of a quantum network \cite{kimble2008quantum,wehner2018quantum,stolk2024metropolitan} can take advantage of state transduction via quantum frequency conversion \cite{kumar1990quantum}. In this process, the quantum state of light is preserved, while its wavelength is shifted. Specifically, difference frequency generation (DFG) enables the compatibility of the zero-phonon emission line of NV centers at $637.2$ nm with low-loss transmission networks and integrated technologies operating in the telecommunication bands \cite{ikuta2014frequency,dreau2018quantum}. Combined with the incorporation of nanodiamonds in fibers \cite{schroder2011fiber}, photonic crystals \cite{faraon2012coupling} and waveguides \cite{faraon2013quantum}, low-noise frequency converters will play a fundamental role in the development of hybrid integrated photonic devices \cite{elshaari2020hybrid,kim2020hybrid}.

Mediated by second-order nonlinearity ($\chi^{(2)}$), DFG is a three-wave mixing process that converts signal photons of frequency $\omega_s$ into a target frequency $\omega_t$.  Respecting energy conservation, the third frequency (pump) is fixed at $\omega_p = \omega_s - \omega_t$. Additionally, the phase-matching condition $\vec{k}_p = \vec{k}_s - \vec{k}_t$ must be satisfied, where $\vec{k}_i, \; i=\{s,t,p\}$ is the wave vector. Such a condition is not trivially reached, but quasi-phase matching can be guaranteed for high efficiency conversion by periodic poling techniques \cite{armstrong1962interactions,franken1963optical}. 

Periodic poled lithium niobate (PPLN) crystals and waveguides are widely used in $\chi^{(2)}$ nonlinear optics applications, including photon sources \cite{konig2005efficient,bock2016highly}, generation of squeezed light \cite{feng2008generation,kashiwazaki2020continuous} and quantum frequency converters \cite{mann2023low,walker2018long,krutyanskiy2019light,meraner2020indistinguishable,de2012quantum,kambs2016low,hannegan2021c,saha2023low,weber2019two,kroh2017heralded,lenhard2017coherence,dreau2018quantum,schafer2023two,bersin2024telecom,ikuta2014frequency}. This photonic platform is particularly promising due to its strong nonlinear coefficient, wide optical transparency window, large refractive index, and high electro-optical effect \cite{qi2020integrated}. Waveguide structures have the additional advantage of being compatible with integrated devices \cite{elshaari2020hybrid,kim2020hybrid}. The preservation of non-classical states after DFG conversion in PPLN waveguides was demonstrated for photons generated on different platforms, such as ion traps \cite{walker2018long,krutyanskiy2019light,meraner2020indistinguishable}, quantum dots \cite{de2012quantum,kambs2016low,weber2019two,hannegan2021c,saha2023low}, devices based on spontaneous parametric down-conversion (SPDC) \cite{kroh2017heralded,lenhard2017coherence} and defect centers in diamonds \cite{dreau2018quantum,schafer2023two,bersin2024telecom}. These results compile successful conversion to target frequencies in the telecommunication O-, C-, and L-bands.

In general, intense pump fields are necessary to efficiently convert photons. In this strong pump regime, excessive background noise due to spontaneous effects, mainly SPDC \cite{pelc2010influence,takesue2010single} and Raman scattering \cite{langrock2005highly,kuo2013reducing,zaske2011efficient,zaske2012visible,pelc2011long}, increases the noise spectral density at target frequencies. A possible strategy to overcome this issue relies on cascaded conversion. By introducing intermediate converted colors, the pump can be highly separated from the target wavelength. This is a viable approach for bypassing the detrimental effects of background noise photons \cite{esfandyarpour2018cascaded}. Two-step DFG was successfully employed for the low-noise conversion of $650$ nm photons to $1588$ nm, within the telecom L-band, in a dual-pumped integrated device \cite{esfandyarpour2018cascaded}. Furthermore, photons from trapped ions \cite{hannegan2021c,saha2023low} and silicon vacancy centers \cite{schafer2023two} were converted to telecom bands using two separated waveguide systems for two-step conversion.

Here, we demonstrate the conversion of $637.2$ nm signals into the telecommunication C-band using a two-stage DFG process in a highly integrated device, i.e., a single-pumped two-section PPLN waveguide. We successfully convert fluorescence photons from NV centers in diamond to the telecommunication C-band with an external (internal) conversion efficiency of $3.0\pm0.1$ ($20.5\pm0.8$) $\%$ and a background noise rate of $2.4\pm0.8$ ($16\pm5$) cps/GHz. The overall losses in our system precluded the operation of the device at single-photon levels. We note that a similar system was recently reported for the conversion of $637$ nm light to the L-band with equivalent noise levels \cite{elsen2024dual}.

\section{Experimental Methods}

Our frequency conversion device consists of an NTT Electronics PPLN on lithium tantalate ridge waveguide with two different periodic poling sections. Each domain of the $40$ mm long waveguide is equally divided in $20$ mm segments and both facets are anti-reflective coated in an effort to minimize coupling losses. In order to avoid photorefractive damage at room temperature, the lithium niobate is doped with 5 mol\% ZnO \cite{bryan1984increased}. The device is glued on a copper tellurium base and mounted in 2 copper casings, where the temperature of each section is actively stabilized with Peltier elements localized underneath them.

\begin{figure}[ht!]
    \centering
    \includegraphics[width=.85\textwidth]{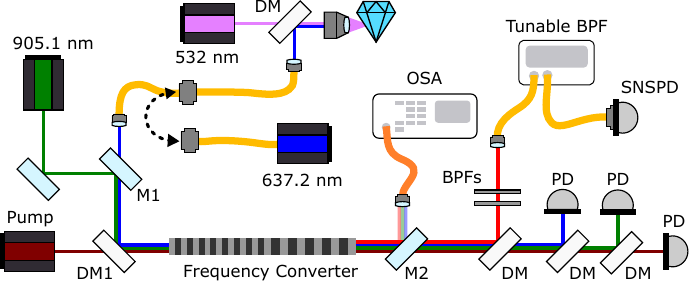}
    \caption{Simplified optical setup. The signal can be taken as either a tunable laser centered at $637.2$ nm or the fluorescence of a cluster of NV centers in diamond. $637.2$ nm or $905.1$ nm are independently combined with the pump laser and send to the conversion system. After the frequency converter, the output fields are either sent to an optical spectrum analyzer (OSA) or are separated for individual detection. M: movable mirror, DM: dichroic mirror, BPF: bandpass filter, PD: photodiode, SNSPD: superconducting nanowire single photon detector.}
    \label{fig_setup}
\end{figure}

The experimental setup is summarized in figure \ref{fig_setup}. The pump field is generated by a Cr${}^{2+}$:ZnS laser, tunable between $1980$ nm and $2528$ nm. Signal and pump are combined with the dichroic mirror (DM$1$) and coupled to the two-step conversion PPLN waveguide with the aid of an uncoated aspheric lens. The signal fiber output provides light emitted either from a reference laser or from a cluster of NV centers in diamond. The movable mirror M$1$ allows the coupling of an auxiliary signal at $905.1$ nm used to evaluate the second-step conversion internal efficiency. Pump, signal and auxiliary signal are coupled to the waveguide with respective efficiencies of $74.5 \; \%$, $88.2 \; \%$ and $81.8 \; \%$.

A second uncoated aspheric lens collimates the out-coupled light into free space, where the four fields involved in the two-step DFG are separated with a combination of dichroic mirrors (DMs). Photodiodes (PDs) are used to monitor the pump, signal and intermediary conversion step fields. The target wavelength is sent to a superconductive nanowire single photon detector (SNSPD), allowing the detection of low-intensity conversion. Stacked free-space bandpass filters (BPFs) reduce the contribution of pump photons to the detection background noise. The tunable bandpass filter has an adjustable bandwidth as narrow as $32$ pm and ranges from $1480$ nm to $1620$ nm, enabling the spectral characterization of the converted light through almost the entire telecommunication band. The movable mirror M$2$ is used for coupling the waveguide output into an optical spectrum analyzer (OSA) (Yokogawa AQ6370D). 

\begin{figure}[ht!]
    \centering
    \includegraphics[width=.85\textwidth]{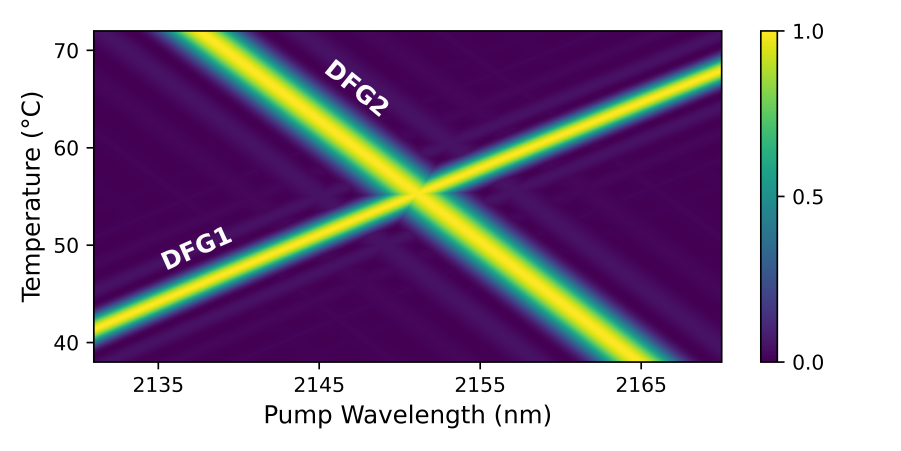}
    \caption{Phase-matching tuning of the two-step conversion. The conversion efficiency heatmaps of the first- (DFG$1$) and second-step (DGF$2$) conversion are shown as a function of each step temperature and the pump wavelength.}
    \label{fig_temptuning}
\end{figure}

Optimal phase-matching conditions in the frequency converter can be achieved by tuning the temperatures of the Peltier elements on each section of the waveguide. Figure \ref{fig_temptuning} shows the calculated phase-matching condition for both steps at different temperatures. The reference values for the theoretical plot followed the Sellmeier coefficients for lithium niobate and lithium tantalate, as provided in \cite{jundt1997temperature} and \cite{barboza2009improved}, respectively. Dispersion effects were considered, where finite-difference time-domain simulations were used to determine the effective refractive index of the propagating fields.

\section{\label{results} Results and Discussion}

In order to avoid detrimental effects on the conversion efficiency due to temperature gradients, we tuned our experiment to a configuration where the first- and second-step conversions are phase-matched to the same temperature. After fine-tuning the pump wavelength to $2152.9$ nm and the crystal temperature to $59.26$ ${}^{\textrm{o}}$C, we successfully converted $637.2$ nm photons to $905.1$ nm followed by a conversion to $1561.6$ nm. This condition slightly differs from the expected degenerate temperature phase-matching conditions of figure \ref{fig_temptuning} due to experimental variations, such as temperature gradients between the waveguide and the temperature sensors, deviations from the calculated refractive indexes, surface roughness and nonparallel walls in our waveguide system.

\begin{figure}[ht!]
    \centering
    \includegraphics[width=.85\textwidth]{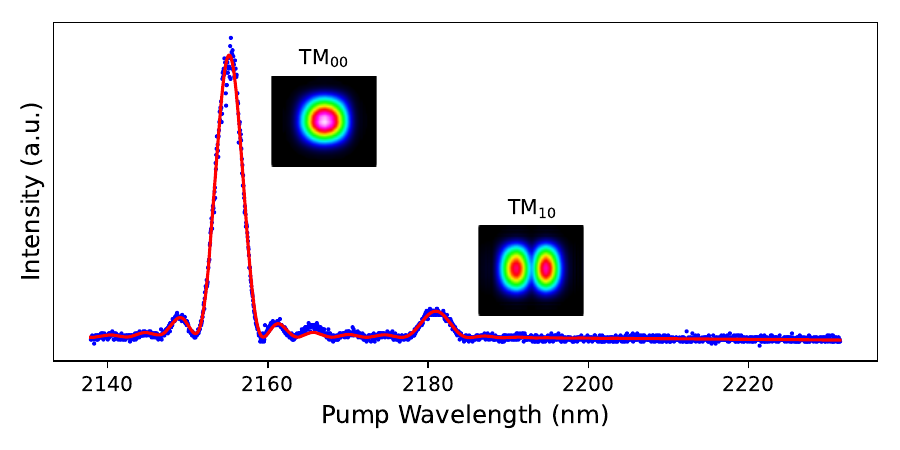}
    \caption{Intensity of the generated light in the first-step conversion to $905.1$ nm as a function of the pump wavelength. The indicated modes were verified with a beamprofiler and are shown in the graphic inset.}
    \label{fig_sinctm00}
\end{figure}

Optimal mode matching was ensured by measuring the transverse profile of the coupled beams after free space out-coupling and by evaluating the first-step conversion of a strong signal field with the aid of a photodetector. Figure \ref{fig_sinctm00} shows the expected conversion proportional to $\textrm{sinc}^2\left( \Delta k L/2 \right)$, where the phase mismatch ($\Delta k$) was tuned with the pump wavelength and $L$ is the length of the phase-matched section of the waveguide respective to the first DFG. A remaining conversion of higher order mode is still visible and the profiles of the modes are depicted in figure \ref{fig_sinctm00} inset.

\begin{figure}[ht!]
    \centering
    \includegraphics[width=.85\textwidth]{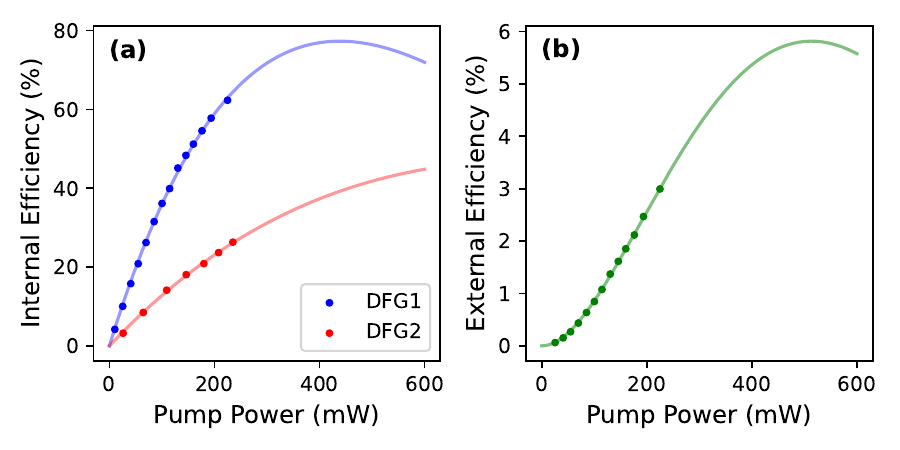}
    \caption{Conversion efficiencies of the two-step converter. DFG$1$ is the conversion from $637.2$ nm to $905.1$ nm and DFG$2$ is the conversion from $905.1$ nm to $1561.6$ nm. (a) Internal conversion efficiencies of the individual steps measured after the waveguide out-coupling. (b) External conversion efficiency of the cascaded process.}
    \label{fig_conversionefficiency}
\end{figure}

The first-step internal conversion efficiency was determined by monitoring the depletion of the $637.2$ nm signal after the waveguide out-coupling. For the investigation of the second-step, we used the auxiliary laser tuned to $905.1$ nm. which corresponds to the wavelength of the conversion of $637.2$ nm light after the first-step conversion. As the auxiliary frequency is also phase-matched to the first-step conversion, sum frequency generation (SFG) is expected, which jeopardizes a depletion measurement. However, because SFG has a narrower phase-matching region than DFG, we were able to suppress it with a small shift on the pump wavelength at the cost of underestimating the second-step conversion efficiency. Figure \ref{fig_conversionefficiency} (a) shows the internal conversion efficiencies as a function of the power of the out-coupled pump. The data is fitted to the equation $\eta_{int} \propto \sin^2\left( \sqrt{\eta_{nor}P_p} L \right)$ \cite{zaske2011efficient}, where $\eta_{nor}$ is the normalized efficiency (in unit per Watt per millimeter squared). The curves indicate our limitation in the coupled pump power, which hinders higher conversion efficiencies. As maximum conversion cannot be achieved in either of the conversion processes, these values cannot be precisely determined.

The total efficiency fitted to the product of the conversion functions is shown in figure \ref{fig_conversionefficiency} (b), reaching an external conversion efficiency limited to $3.0 \pm 0.1  \; \%$ corresponding to an out-coupled pump power of $225$ mW. This value was reached by comparing the total photon number of a strong signal coupled into the waveguide with the number of converted photons after our filtering system. The power measurements were performed with the aid of a power meter. In order to isolate the converted photons from remaining pump photons and any detrimental background noise, we used stacked BPFs centered at 1550 nm with $40$ nm FWHM followed by a $32$ pm tunable filter centered at the target wavelength of $1561.6$ nm. The total losses at this wavelength sum up to $85.4 \pm 0.1 \; \%$ and have contributions of the out-coupling efficiency ($92.2\pm0.2 \; \%$), free-space filters and optical elements transmission ($94.7\pm0.2 \; \%$), SMF coupling efficiency ($82.6\pm0.1  \; \%$) and tunable BPF transmission ($20.2\pm0.1 \; \%$). Accounting for the losses, we obtain a maximum internal conversion of $20.5 \pm 0.8 \; \%$, limited by the accessible pump power.

As a next step we investigated the tuning capability of our device. Following the phase-matching conditions of figure \ref{fig_temptuning} we were able to tune the target wavelength by shifting the temperature of the second poling section of the converter away from the temperature degeneracy point. With a gradient of $\Delta T = 4. 6$ ${}^{\textrm{o}}$C we reached a target wavelength of $1559.0$ nm with an external conversion efficiency of $2.8\pm0.3 \; \%$. Furthermore, by setting $\Delta T = -6.1$ ${}^{\textrm{o}}$C, the reached configuration was $1564.9$ nm with $2.9\pm0.2 \; \%$ of efficiency. Despite the temperature gradients induced between the poled sections the conversion efficiency remained within the error margins in this tuning range. The tunability limits were not explored in order to avoid damages to the chip due to physical stress.

\begin{figure}[ht!]
    \centering
    \includegraphics[width=.85\textwidth]{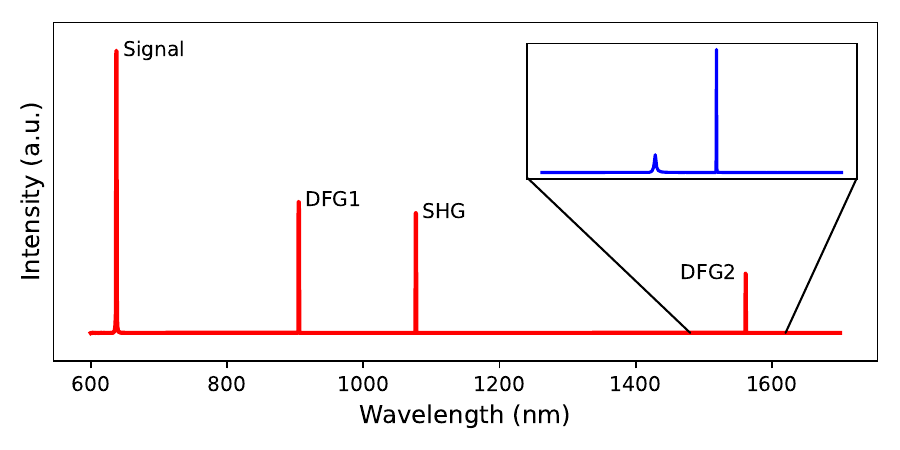}
    \caption{Optical spectrum of the converter output. First (DFG$1$) and second (DFG$2$) frequency difference generation steps are indicated in the figure, as well as the second harmonic generation (SHG) of the pump field. The inset shows the spectral region from $1480$ nm to $1620$ nm, measured with the combination of the tunable BPF with the SNSPD, unraveling an extra peak near DFG2.}
    \label{fig_OSA}
\end{figure}

The overall spectrum of the device output shown in figure \ref{fig_OSA} was measured with an OSA, where we verified all the involved fields in our conversion process within the equipment range. The peak at $1076.5$ nm is due to second harmonic generation of the pump in the waveguide, where its expected quadratic response with the pump power \cite{fejer1992quasi} was experimentally verified. 

We thoroughly investigated the target spectrum around the wavelength of $1561.6$ nm (inset in figure \ref{fig_OSA}) using the combination of the tunable BPF and an SNSPD. Noise near the target wavelength is a critical limitation of frequency converters. Therefore, we characterized the noise spectral density of our system by setting the tunable BPF to the target wavelength with minimum bandwidth ($4$ GHz). Taking an integration time of $20$ minutes, we reached a total count rate of $142 \pm 1$ cps, of which $ 135\pm2$ cps are attributed to dark counts. Accounting for the
SNSPD detection efficiency of $72\pm3 \; \%$, we reach a pump induced noise count rate of $10\pm3$ cps, corresponding to an external noise spectral density of $2.4\pm0.8$ cps/GHz. Correcting for the system external losses, we obtain an internal noise spectral density of $16\pm5$ cps/GHz. The background noise is comparable to that of other reported two-step conversion schemes \cite{esfandyarpour2018cascaded,schafer2023two,elsen2024dual,saha2023low}, while the device itself is more compactly designed.

A prominent feature near the target wavelength is a broad peak centered at $1532.8$ nm. Further investigations into the source of this contribution to noise led to a process occurring in the waveguide independently from the signal field. This is a phase-matched conversion related to the second-step periodic poling, where its central wavelength shifted linearly by $-0.045$ nm/ºC within a $10$ ºC temperature detuning range. Its frequency also scales linearly with the pump wavelength, remaining at a constant separation of $56.34\pm0.01$ THz. The efficiency of the generated peak is highly dependent on the polarization of the pump, further indicating its relation to the periodic poling structure. Such temperature and strong polarization dependence discards the possibility of a pure Raman scattering origin. Finally, the process showed a linear dependence with the pump intensity, eliminating any wave mixing processes between different pump harmonics. The origin of this peak is then attributed to SFG of the pump combined with thermal photons in the range of $5.3 \; \mu$m.

\begin{figure}[ht!]
    \centering
    \includegraphics[width=.85\textwidth]{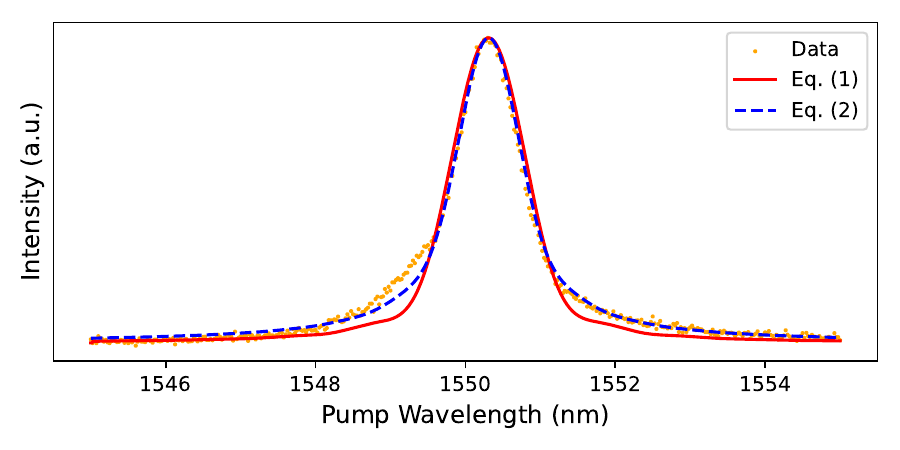}
    \caption{Noise peak caused by thermal photons converted by SFG. The different fittings are given by the analytical model of Eq. (\ref{eq_noise}) and the numerical modification of Eq. (\ref{eq_noise2}).}
    \label{fig_noisepeak}
\end{figure}

We were able to confirm this speculation by a theoretical model. The detailed line shape of the converted thermal photons is shown in figure \ref{fig_noisepeak}. Note that we displaced its center wavelength from $1532.8$ nm to the center of the free-space BPFs at $1550$ nm in order to avoid any distortions caused by the filter cutoff.
Assuming that the thermal emission can be described by black body radiation, the mid-infrared spectrum remains flat. Hence, an underlying phase-matching curve of the form $\textrm{sinc}^2\left( \Delta k L/2 \right)$ is expected. Considering, however, that the thermal photons are generated at different positions along the waveguide, the total output intensity will be given by \cite{pedersen2014non}
\begin{equation}
\begin{split}
    I &\propto \int^L_0 \frac{\left( L - z \right)^2}{L} \textrm{sinc}^2 \left( \frac{\Delta k (L-z)}{2} \right) \textrm{d}z \\
    &= \frac{2}{\Delta k^2}\left[1 - \textrm{sinc}(\Delta kL) \right].
    \label{eq_noise}
\end{split}
\end{equation}
Note that we did not consider any contribution from the first half of the waveguide to the thermal photons, as lithium niobate is highly absorptive for this frequency range \cite{leidinger2015comparative}. The fitting of the analytical model is shown in figure \ref{fig_noisepeak}, which is in good agreement with the data. The deviation in the tails of the function is due to the assumption of a uniform distribution of thermal photons along the waveguide. A better description of the noise peak is numerically obtained by including weighting parameters along the waveguide length according to
\begin{equation}
    I \propto \sum_{z} \sum_{i=0}^n a_i z^i \textrm{sinc}^2\left(\frac{\Delta k z}{2}\right),
    \label{eq_noise2}
\end{equation}
where $z \in [0,L]$, $a_i$ is the free parameter and we chose $n=2$ to avoid over fitting. The asymmetry of the peak is likely due to non-uniformities in the waveguide.

\begin{figure}[ht!]
    \centering
    \includegraphics[width=.85\textwidth]{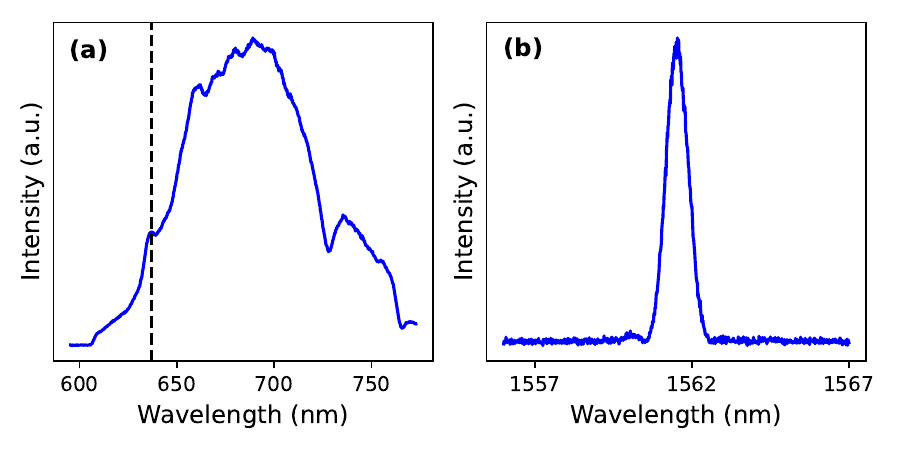}
    \caption{Cascaded conversion of fluorescence light from NV centers in diamond. (a) Fluorescence spectrum of the optical source. The vertical line indicates the zero-phonon line. (b) Converted spectrum from the surroundings of the zero-phonon line to telecommunication C-band frequencies.}
    \label{fig_diamondconversion}
\end{figure}

A targeted task of our converter is to convert photons from NV centers to the telecom C-band. Therefore, as a final demonstration, we converted fluorescence photons collected from a cluster of NV centers in diamond with a confocal microscope configuration \cite{schroder2011ultrabright}, as represented in figure \ref{fig_setup}. We off-resonantly excite the defect centers with a $532$ nm pump and connect the collected photons to the converter with a SMF. The spectrum of the color centers emission is shown in figure \ref{fig_diamondconversion} (a), where the zero-phonon line is indicated at $637.2$ nm. The resulting spectrum after the cascaded conversion process is shown in figure \ref{fig_diamondconversion} (b). The characteristic $\textrm{sinc}^2$ function can be faintly seen after conversion, as the fluorescence spectrum of the bulk diamond is much broader than the conversion bandwidth.

Given the efficiency of our two-step conversion and the overall loss budget from the confocal microscope to the frequency converter device, a significant number of NV centers contributed to the fluorescence signal, resulting in approximately $10$ $\mu$W of optical power. Nonetheless, the frequency conversion of non-classical states in similar compact devices would benefit from the achieved low-noise performance. Further improvements in conversion efficiency, attainable through asymmetric poling lengths, would enable the conversion of signals in the single-photon regime. Additionally, the conversion of thermal photons can be suppressed by redesigning the poling periods. An elaborated discussion on the limits of single-photon frequency conversion can be found in reference \cite{ikuta2014frequency}.

\section{\label{conclusion} Conclusion}

We experimentally demonstrated the two-step conversion of light emitted by NV centers in diamond to the telecommunication C-band, where we used a single-pumped waveguide device. The tunability of the system was demonstrated by changing the phase-matching conditions with temperature control of the individual DFG steps. This feature can be further optimized with thermally isolated poled sections and can be precisely controlled with the aid of localized microheaters \cite{lee2017chip}. We observed a total external (internal) conversion efficiency of $3.0\pm0.1$ ($20.5\pm0.8$)$\%$. Higher efficiencies can be reached with asymmetric poling lengths, individually optimized to each conversion step. Our cascaded  two-step conversion in a single waveguide not only provides a highly integrated approach, but at the same time it reduces the background noise from Raman scattering and SPDC. We observed an external (internal) noise spectral density of $2.4\pm0.8$ ($16\pm5$) cps/GHz, which is on par with other reported quantum converters. The low background noise can be attributed to remaining anti-Stokes Raman scattering of the pump \cite{pelc2011long}. Surprisingly, phase-matched SFG of thermal photons provided a significant contribution to the noise as we investigated in detail experimentally and theoretically. This can be bypassed by choosing different poling periods, however it is important to take this extra noise source into account when designing future integrated conversion devices.

\begin{backmatter}
\bmsection{Funding}
This project was funded by BMBF, QR.X  project 16KISQ003. R.A.K. acknowledges funding by the Senat Berlin.

\bmsection{Acknowledgment}
We thank Tim Kroh for his contributions to early implementations of the project and Sven Ramelow and Felix Mann for valuable discussions and insights on the converted spectrum.

\bmsection{Disclosures}
The authors declare no conflicts of interest.

\bmsection{Data Availability Statement}
Data underlying the results presented in this paper are not publicly available at this time but may be obtained from the authors upon reasonable request.

\end{backmatter}

\bibliography{sample}

\end{document}